# Sequential Strangeness Freeze-out

*Rene* Bellwied[1,*]

[1]University of Houston, Physics Department, 617 SR1 Bldg., Houston, TX 77204, USA

**Abstract.** I will describe the latest results from lattice QCD pertaining to a potential flavour hierarchy in the hadronic freeze-out from the QCD crossover region. I will compare these results to a variety of improved hadronic resonance gas calculations and to experimental data of fluctuations of net-charge, net-proton and net-kaon multiplicity distributions, which serve as a proxy for the susceptibilities of conserved quantum numbers on the lattice. I will conclude that there is intriguing evidence for a flavour dependent freeze-out, and I will suggest expansions to the experimental program at RHIC and the LHC that could potentially demonstrate the impact of a flavour separation during hadronization.

## 1 Introduction

The detailed determination of a pseudo-critical temperature based on continuum extrapolations of the temperature dependence of the chiral susceptibility on the lattice, in comparison to calculations of the chemical freeze-out temperature using particle yields at RHIC and the LHC, seems to indicate that hadronization and freeze-out coincide near the phase boundary in the QCD phase diagram. The question arises whether this transition from quark to hadron degrees of freedom occurs at the same temperature for all particle species and/or quark flavours. The application of statistical hadronization models is successful in describing hadronic particle yields over many orders of magnitude. From the abundant pion yields to the rare alpha particle, the thermally equilibrated system can be broadly defined with two common freeze-out parameters, namely the chemical freeze-out temperature and the baryo-chemical potential. These calculations were applied over a wide range of collision energies from the SPS to the LHC, and system sizes from pPb to PbPb. However, recent high resolution measurements of the particle yields in ALICE at the LHC and STAR at RHIC, as well as the net-particle fluctuations in STAR, seem to indicate that there might be evidence for a sub-structure in the common freeze-out picture. The early LHC measurements attributed the tension in a common fit to a 'proton anomaly', because, at the time, only the proton yields seemed to deviate from the anticipated particle yields. As of late, though, i.e. with the Run-2 data from ALICE, also the multi-strange $\Xi$ baryons show a significant deviation from the common temperature fit. These results can be related to attempts to deduce the freeze-out temperatures of particular conserved quantum numbers independently from fluctuations of net-particle distributions and lattice QCD.

[*] Corresponding author: bellwied@uh.edu

## 2 Susceptibilities from lattice QCD

Susceptibilities are defined as the derivatives of the pressure with respect to the chemical potential. Continuum extrapolated susceptibility calculations of single flavour quantum numbers showed that there is a difference between flavours in the crossover region [1], see Fig.1(left). Fig.1(right) shows the flavour specific susceptibility ratio $\chi_4/\chi_2$ [2], which was suggested as a specific observable to deduce chemical freeze-out temperatures directly [3], from a comparison of experimental data to first principle calculations. The lattice data themselves show a peak at different temperatures and their agreement with Hadron Resonance Gas (HRG) model calculations begins to deviate at these temperatures as well. This is, without a direct comparison to experimental data, not yet proof of a flavour hierarchy in the crossover region, but it is suggestive of different freeze-out temperatures for light and strange flavour particles.

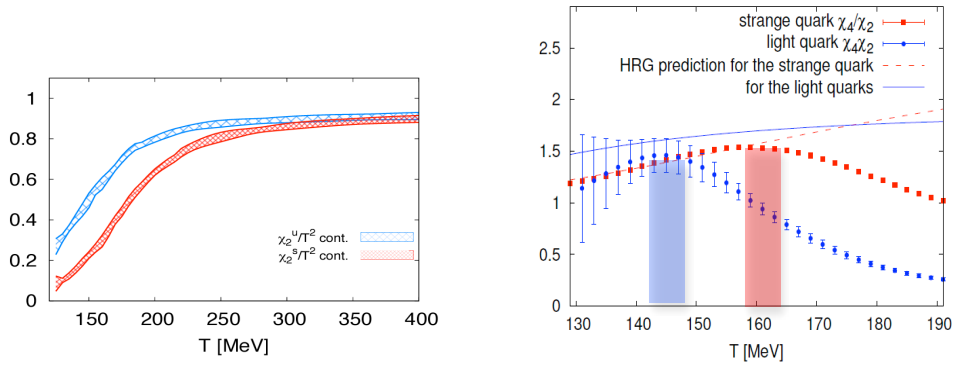

**Fig. 1.** (left): Continuum extrapolated lattice QCD results for $\chi_2^u$ and $\chi_2^s$ [1], (right): Continuum extrapolated lattice QCD results for $\chi_4/\chi_2$ for light and strange quarks in comparison to HRG model calculations [2].

## 3 Comparison to the Hadron Resonance Gas Model

The relevance of the hadronic spectrum included in the HRG calculations was highlighted early on by Bazavov et al. [4] in order to explain a potential deviation from the lattice curves at different temperatures for different flavours. It was shown that the inclusion of certain states significantly improves the agreement between HRG and lattice QCD for certain susceptibilities in the QCD crossover region. The initial study simply expanded the list of available states by expectations from the non-relativistic Quark Model [5]. More refined Quark Model variations, which include the possibility of quark-quark interactions in the hadron, significantly reduce the number of expected states, while at the same time the number of experimentally verified states in the listing of the particle data group, e.g. PDG-2016 [6], continues to increase. A detailed study of the HRG approach including the latest findings, was presented at this conference and in a recent publication [7, 8] and indicated that a compromise between too many and too few excited states in the high mass hadronic spectrum can be found, based on all states listed in PDG-2016. With this input the HRG calculation can describe most features of a multitude of susceptibility ratios calculated with lattice QCD up to the pseudo-critical temperature(s). As an example, Fig.2 shows the two most relevant susceptibility ratios that can be measured to determine flavour dependencies, namely the $(\chi_S/\chi_B)_{LO}$ and the $\chi_4/\chi_2$ for strange quarks [8]. It is evident that the so-called PDG-2016+ list gives the best compromise between old PDG and QM listings.

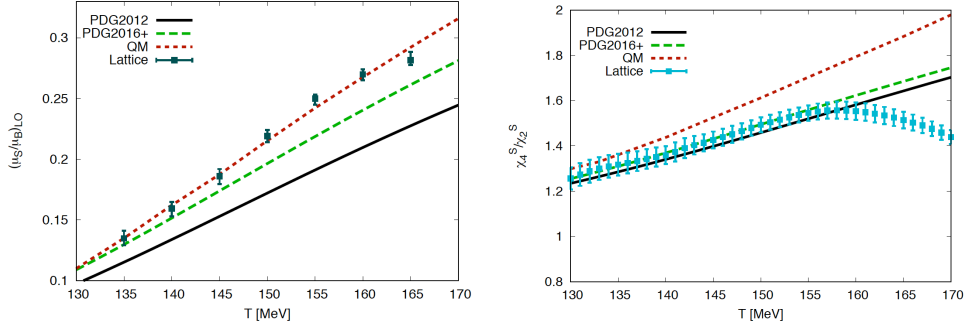

**Fig. 2.** Continuum extrapolated lattice QCD results for $(\chi_S/\chi_B)_{LO}$ (left) and for $\chi_4^s/\chi_2^s$ (right) in comparison to HRG model calculations with varying number of resonant states based on PDG-2012, PDG-2016+ (i.e. incl. one star states), and non-relativistic Quark Model predictions [8].

A more provocative extension of the standard HRG approach was also presented at this conference and in a recent publication, expanding on the question whether a non-interacting resonance gas is indeed the most realistic proxy for the hadronic interaction strength near the phase transition [9]. As an alternative the authors proposed to either use a parametrization of the low energy van-der-Waals interactions or an excluded volume to describe the attractive and repulsive hadron-hadron interactions. These changes to the HRG approach seemingly extend the agreement between HRG and lattice QCD to higher temperatures, but a.) the agreement now reaches beyond the pseudo-critical temperature and b.) the rather unconstrained parameter base for the interactions and the excluded volume allows for rather large variations in the fit. Nevertheless the approach is intriguing and, in the future, could also take into account potential flavour dependent differences in the parameters.

## 4 Fits based on experimental studies of fluctuations and yields

Recent results published by the STAR collaboration [10] regarding the evolution of the chemical freeze-out temperature for the various energies of the RHIC beam energy scan showed that although a common freeze-out temperature can be found through fits to the measured yields of all particles, this temperature will be about 15-20 MeV lower, if only light flavour particles (and kaons) are included in the fit (see Fig. 35 in [10]). The kaon yield is included, but was shown to be rather insensitive to the freeze-out temperature [11]. Independently the work by Chatterjee et al. [12] showed that the fits to ALICE yields significantly improve if two separate chemical freeze-out temperatures are assumed for light and strange particles. This requirement for two temperatures seemingly disappears when the yields in small systems (pp and pPb collisions) are fitted [13].

Regarding the use of net-particle fluctuations rather than the yields, our studies that tried to determine net-electric charge and net-baryon number fluctuations, by using the net-charged particle and net-proton distributions as a proxy, found a chemical freeze-out temperature for these particles, based on a HRG fit, that was consistently 15-20 MeV below the expected value from a common fit to all measured particle yields, see Fig.3(left) [14].

The resulting temperatures are in agreement with lattice QCD calculations based on the related susceptibility ratios [15,16]. One should note that all HRG fits are applied to the $\chi_2/\chi_1$ ratios for a particular particle species. We found that the error bars for the higher moments are still too large and that the higher moments might potentially have contributions from critical fluctuations at the RHIC beam energies, which will negatively impact the chemical freeze-out fits [14]. In our conclusions, we pointed out that both, the net-charged particles and the net-protons, are dominated by light quark particles (pions and protons, respectively), and that a fit to a strangeness proxy, i.e. the fluctuations of net-kaons, might shed light on the question whether the chemical freeze-out could indeed be flavour dependent. In Fig.3(right) I am presenting preliminary results of a HRG fit to the recently published net-kaon data from STAR [17]. The HRG model is the same used in Fig.3(left), in fact the baryo-chemical potential was adopted from the fit to net-protons and net-charges and only the temperature was left as a free parameters. As one can see the net-kaons tend towards higher freeze-out temperatures than the other net-particle distributions. This might be surprising because as mentioned before, the kaon yields are rather insensitive to the temperature in a statistical hadronization fit, but our group also showed that the higher order moments of the net-kaon distributions are more sensitive to the freeze-out parameters than the yields themselves [18]. In Fig.3(right) we also show that the final result is rather insensitive to the number of hadronic states included in the HRG calculation. Lattice calculations assuming a Boltzmann approach to the partial pressure in order to isolate the contribution of the kaons to the strangeness susceptibilities confirm this result [19].

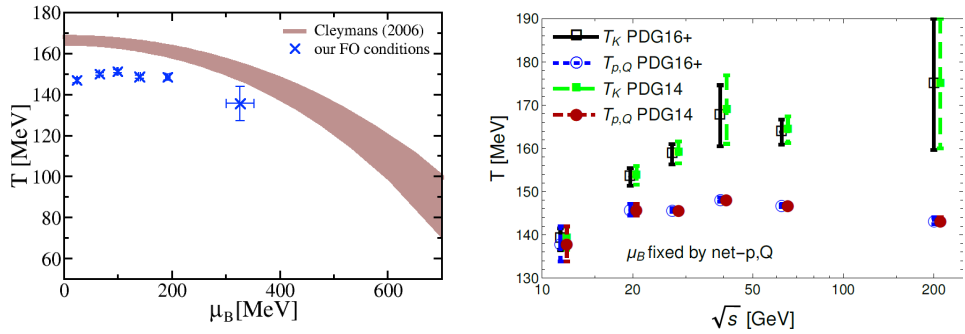

**Fig. 3.** (left): Results from a combined HRG fit to the $\chi_2/\chi_1$ measurements for net-charges and net-protons from STAR (blue points) [14]. The extracted freeze-out parameters are compared to the curve by Cleymans et al. [20], which tried to parametrize all freeze-out results from statistical hadronization models to yields from SIS, SPS, RHIC and LHC. (right): Preliminary results from fitting the $\chi_2/\chi_1$ measurements for net-kaons from STAR with the same model used in Fig.3(left). The $\mu_B$ was fixed by the net-p,Q results. The figure also shows a comparison between results using different PDG lists in the HRG calculation.

One should note that the usage of net-kaons as a proxy for net-strangeness is less justified than the proxies for net-charge and net-baryon number [21]. Certainly the inclusion of fluctuation data for strange baryons is very important in order to gauge the relative contributions of flavour and baryon number to the final result. Studies of these fluctuations are underway in STAR and ALICE.

## 3 Conclusions and Outlook

There is intriguing evidence that the flavour composition of the produced hadrons might play a role in their freeze-out parameters and thus in their hadronization dynamics. Particularly it seems that the quark mass plays a significant role in calculating the transition of flavour specific susceptibilities on the lattice as long as the mass is not negligible compared to the temperature of the equilibrated system. Experimentally this could mean that heavier quark particles prefer to freeze-out at a higher temperature. I have presented evidence to that effect based on yield and fluctuation measurements from STAR and ALICE. Studies of the baryon mass evolution in the crossover region, based on PNJL [22] and lattice calculations [23], also point at finite quark mass, and thus flavour, dependencies. Certainly more evidence is needed, in particular from strange baryon fluctuation measurements, but in theory this flavour dependence should also be measurable in the charm sector as long as the charm quarks thermalize with the system.

A direct impact of a higher freeze-out temperature would be an enhancement of strange particles relative to the yield obtained at a common lower freeze-out temperature. Indirect evidence can be found in the ALICE data based on the latest $\Xi$ yields in central PbPb collisions and on the comparison of PbPb yields to pp yields for strange particles. The latter point is often attributed to canonical suppression in the small system [24], but a comparison between preliminary RHIC-BES data [25] and ALICE data [26] leads us to believe that the energy dependence of the canonical suppression makes this effect almost negligible at LHC energies, see Fig.4. Based on the data it seems that the strange anti-baryon yields follow the same trend from sqrt(s) = 62 GeV on up. In addition, HRG fits, assuming a thermal system is produced in pp collisions, show that the temperature in the small systems is about 15 MeV lower than the strange particle freeze-out temperature in heavy ion collisions [12], which could explain at least part of the strangeness enhancement. Further studies are needed, though.

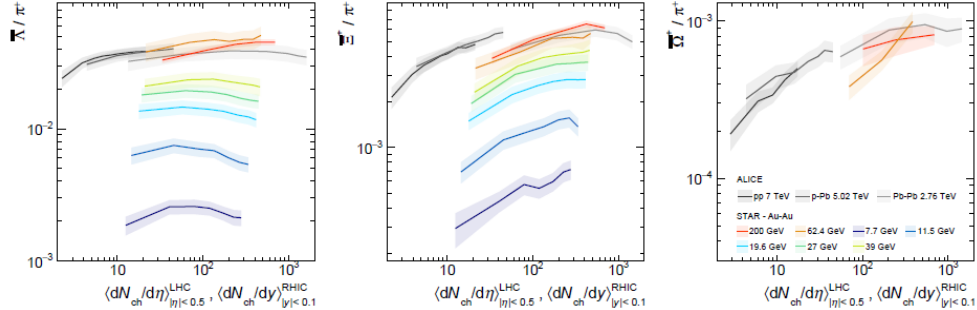

**Fig. 4.** Anti-baryon over p- production as a function of collision energy and charged particle density (for LHC: $\langle dN_{ch}/d\eta \rangle$ in $|\eta|<0.5$, for RHIC: $\langle dN_{ch}/dy \rangle$ in $|y|<0.1$) based on STAR [25] and ALICE [26] data.

Regarding a more speculative aspect of particle production, this enhancement in strange quarks could lead to strangeness clustering, which might manifest itself in strange multi-quark states. The discovery of charmed tetra-and penta-quark states by LHCb certainly has triggered renewed interest in exotica searches in the strangeness sector.

In terms of more dynamic quantities, a higher freeze-out temperature could potentially lead to a shortened partonic phase for strange baryons. This could reduce dynamic quantities, such as $R_{AA}$ and $v_2$, as long as a significant contribution to the flow or the suppression is generated close to pseudo-critical temperature.

## Achnowledgements

I would like to thank my collaborators for various contributions to this work: Paolo Alba, Livio Bianchi, Szabolcs Borsanyi, Zoltan Fodor, Jana Guenther, Anders, Knospe, Valentina Mantovani-Sarti, Jackie Noronha-Hostler, Paolo Parotto, Attila Pasztor, and Claudia Ratti. I also thank Kryzstof Redlich and Frithjof Karsch for some inspiring discussions. This work was supported by the U.S. Department of Energy under DE-FG02-07ER41521.